\documentclass[12pt,a4paper]{article}
\usepackage[dvips]{graphicx}
\begin{document}
\textwidth=135mm
\textheight=200mm

\begin{center}{
{\bfseries {The study of the thermal neutron flux in \\
the deep underground laboratory DULB-4900}}\footnote{
{\small Talk at The International Workshop on Prospects of Particle Physics: "Neutrino
Physics and Astrophysics" February 01 - Ferbuary 08, 2015, Valday, Russia}}

\vskip 5mm

V.V. Alekseenko$^{\dag}$, Yu.M. Gavrilyuk$^{\dag}$, A.M. Gangapshev$^{\dag}$, A.M. Gezhaev$^{\dag}$,
\\D.D. Dzhappuev$^{\dag}$, V.V. Kazalov$^{\dag}$, A.U. Kudzhaev$^{\dag}$, V.V. Kuzminov$^{\dag}$,
\\S.I. Panasenko$^{\ddag}$, S.S. Ratkevich$^{\ddag}$, D.A. Tekueva$^{\dag}$, \\
S.P. Yakimenko$^{\dag}$, Yu.V. Stenâ'kin$^{\dag}$

\vskip 5mm

{\small {\it $^\dag$ Institute for Nuclear Research, RAS, Moscow, Russia}}
\\
{\small {\it $^\ddag$ Kharkiv National University, Kharkiv, Ukraine}}
}
\end{center}

\vskip 5mm

\centerline{\bf Abstract}
We report on the study of thermal neutron flux using monitors based on mixture of ZnS(Ag) and LiF enriched with a lithium-6 isotope at the deep underground laboratory DULB-4900 at the Baksan Neutrino Observatory. An annual modulation of thermal neutron flux in DULB-4900 is observed. Experimental evidences were obtained of correlation between the long-term thermal neutron flux variations and the absolute humidity of the air in laboratory. The amplitude of the modulation exceed 5\%
of total neutron flux.

\vskip 3mm


\section{\label{sec:intro}Introduction}

The characterization of the neutron fluence has become a critical issue for experiments that require these extreme low-background environments, such as $\beta\beta$-decay, dark matter searches, and solar neutrino experiments \cite{PRC13, PEPAN15XE, Gavrin11}. As direct detection dark matter experiments seek to detect WIMP dark matter by observing nuclear recoils produced by WIMP interactions with nuclei, and as single scatter neutron events can produce nuclear recoils that are indistinguishable from WIMP interactions, the issue of in situ background control is particularly relevant in considering neutron backgrounds. Annual modulation of neutron flux in underground facilities could mimic annual modulation of signal from WIMPs, due to elastic scattering on nuclei of target. The neutron flux could depend on parameters such as humidity, temperature, pressure and other meteorological conditions. The long term measurement of thermal neutron flux is performed to study these dependencies for underground facilities at the Baksan Neutrino Observatory of the Institute for Nuclear Researches Russian Academy of Science (BNO INR RAS) \cite{EPJPkuz12}.

\section{Experimental setup}

The measurements are performed in box A1 of deep underground laboratory DULB-4900 (at a depth of 4900 m w.e., cosmic ray flux is reduced by a factor $10^7$) {\cite {DULB}}. Where the temperature of the air is $\sim27^o$~C and is stable during the year. The experimental setup consist of four detectors (D1, D2, D3 and D4) and three sensors (temperature, humidity, pressure). The detectors are identical to that used in {\cite{Stenkin}}. The schematic view of the setup is presented in Fig.~{\ref{fig:scheme}}. Each detector is an aluminium container with sizes  $700\times700\times300$~mm$^3$ which is viewed by photomultiplier tube (FEU-49), see Fig.~{\ref{fig:detector}. Thin scintillator made of solid granulated alloy $^6$LiF+ZnS(Ag) ($\sim 3/1$ in mass, thickness is $\simeq30$~mg/cm$^2$) with the area of $600\times600$~mm$^2$ is placed on the bottom of the container. The reaction $^6$Li+$n \rightarrow ^{3}$H + $\alpha$ is used to detect thermal neutrons (Q=4.79~MeV, $E_\alpha = 2051$ keV, $E_H=2735$ keV, $\sigma = 945$~b at 300~K). The signals from each PMTs are splited after preamplifier, one goes to summator, another goes directly to input of digital oscilloscope. The signal after summator goes to input of digital oscilloscope as well. The sample of recorded signal is shown in Fig.~{\ref{fig:pulse}}.
\begin{figure}[pt]
\begin{center}
\includegraphics[width=2.5 in, angle=0]{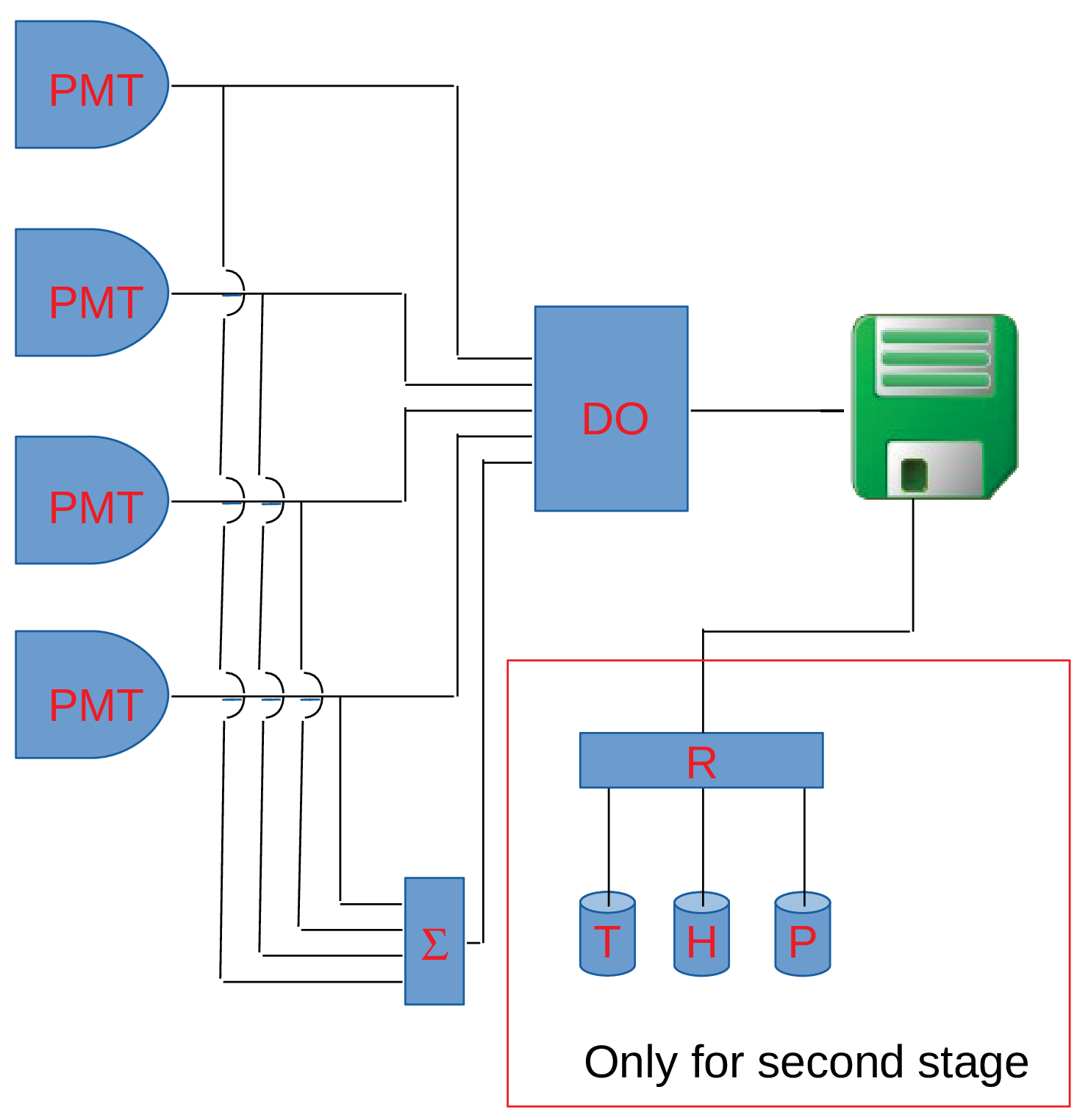}
\caption{\label{fig:scheme} Schematic view of the setup. PMT - photomultiplier, $\Sigma$ - sumator, DO - digital oscilloscope, R - recorder, T - temperature sensor, H - humidity sensor, P - pressure sensor}
\end{center}
\end{figure}
Each detector is an aluminium container with sizes  $700\times700\times300$~mm$^3$ which is viewed by photomultiplier tube (FEU-49), see Fig.~\ref{fig:detector}.
\begin{figure}[pt]
\begin{center}
\includegraphics[width=1.5 in, angle=0]{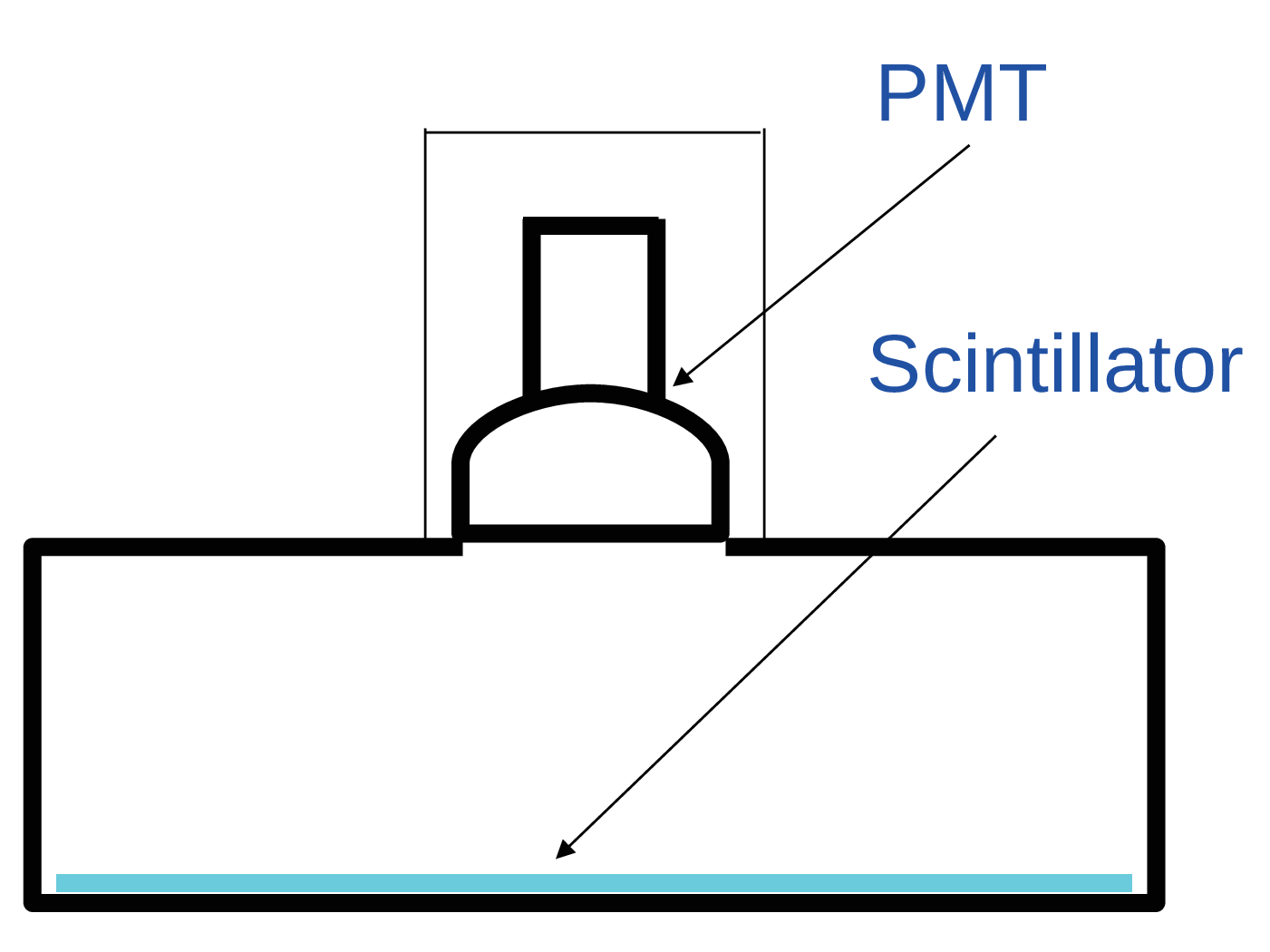}
\caption{\label{fig:detector} Schematic view of the detector.}
\end{center}
\end{figure}
Thin scintillator made of solid granulated alloy $^6$LiF+ZnS(Ag) ($\sim 3/1$ in mass, thickness is $\simeq30$~mg/cm$^2$) with the area of $600\times600$~mm$^2$ is placed on the bottom of the container. Detectors are similar to that proposed in {\cite{Stenkin}}. The reaction $^6$Li+$n \rightarrow ^{3}$H + $\alpha$ is used to detect thermal neutrons (Q=4.79~MeV, $E_\alpha = 2051$ keV, $E_H=2735$ keV, $\sigma = 945$~b at 300~K). The signals from each PMTs are splited after preamplifier, one goes to summator, another goes directly to input of digital oscilloscope. The signal after summator goes to input of digital oscilloscope as well. The sample of recorded signal is shown in Fig.~\ref{fig:pulse}.
\begin{figure}[pt]
\begin{center}
\includegraphics[width=3.0 in, angle=0]{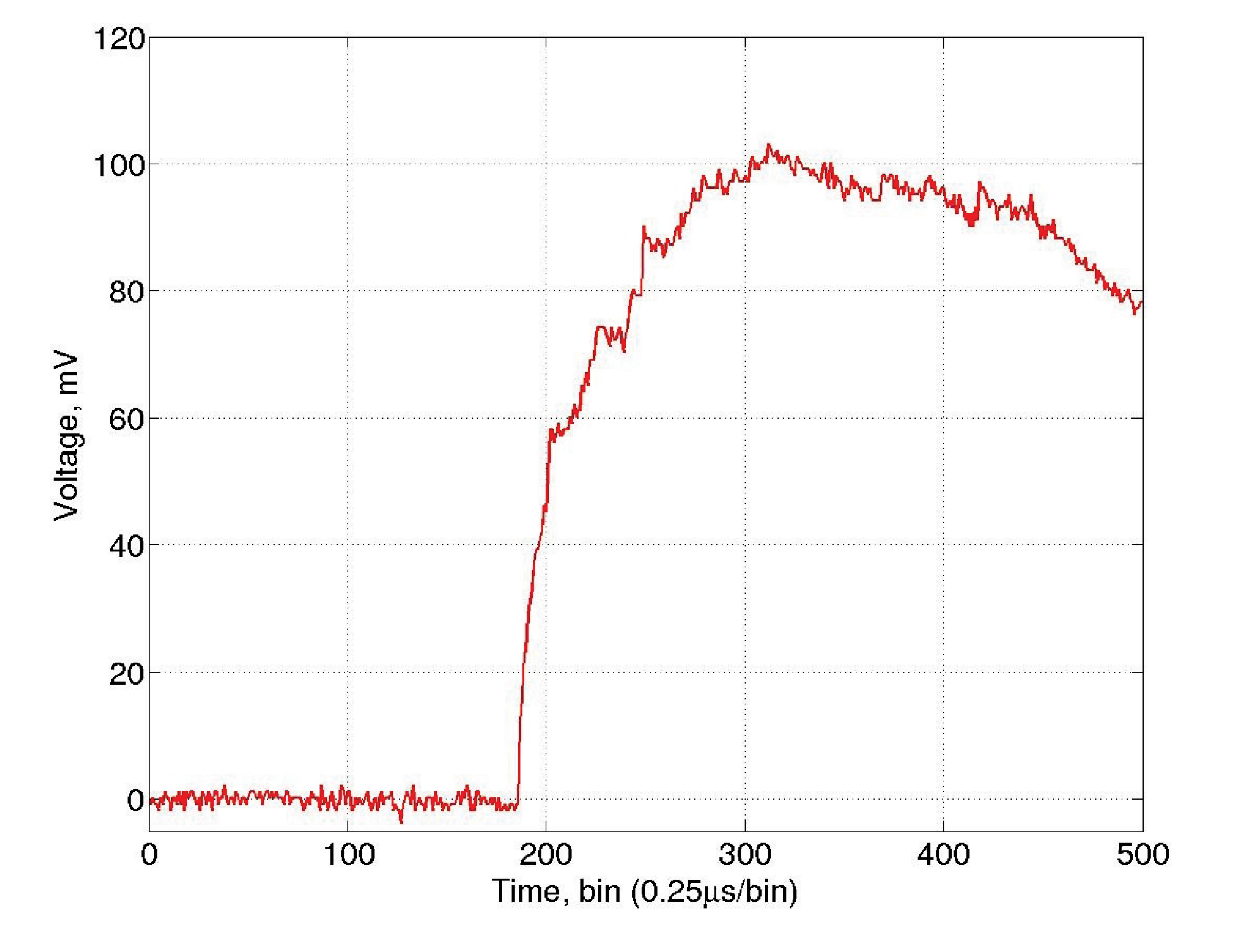}
\caption{\label{fig:pulse} Pulse sample.}
\end{center}
\end{figure}

Between first and second stages, detectors were alternately calibrated with neutron source ($n$-source). $n$-source consisted of set of $\alpha$-sources ($^{238}$Pu + $^{239}$Pu + $^{226}$Ra + triplet / $^{233}$U + $^{238}$Pu + $^{239}$Pu /, total activity - 131~kBq) covered with $0.1$ mm beryllium foil. Sources were placed inside cylindrical polyethylene box. The thickness of the box wall is $5$ cm. Neutrons come from reaction $^9$Be+$\alpha \rightarrow$ $^{12}$C + $n$ (Q=5.5~MeV, $E_n\simeq1\div8$ MeV, $\sigma\simeq0.1$~b, $p_{\alpha,n}\simeq2.5\cdot10^{-5}$). Estimated activity of $n$-source is $A_n\simeq3.28$~s$^{-1}=1.18\cdot10^4$~h$^{-1}$.

The $n$-source was attached to the bottom of detectors (close to the scintillator), so $\sim 50$\% of neutrons cross the scintillator.
But only small part of neutrons are thermalised after passing 5 cm of polyethylene. Spectra collected with detector D1 are presented in Fig.~\ref{fig:spectra} as example.
The response of all detectors on $n$-source are similar, with  increasing of count rate by 700~cts/day (see Fig.~\ref{fig:calibration}).
\begin{figure}[t]
\begin{center}
\includegraphics[width=3.5 in , angle=0]{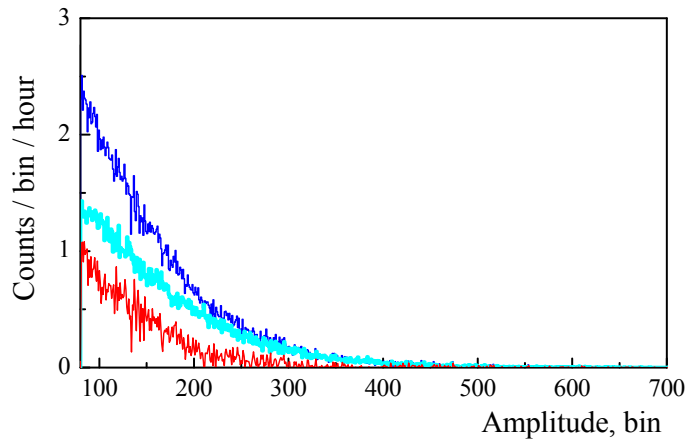}
\caption{\label{fig:spectra} Spectra from detector D1.
Blue thin line - background plus $n$-source, motton blue line - background,
red line - difference by $n$-source.}
\end{center}
\end{figure}
\begin{figure}[t]
\begin{center}
\includegraphics[width=3.5 in, angle=0.]{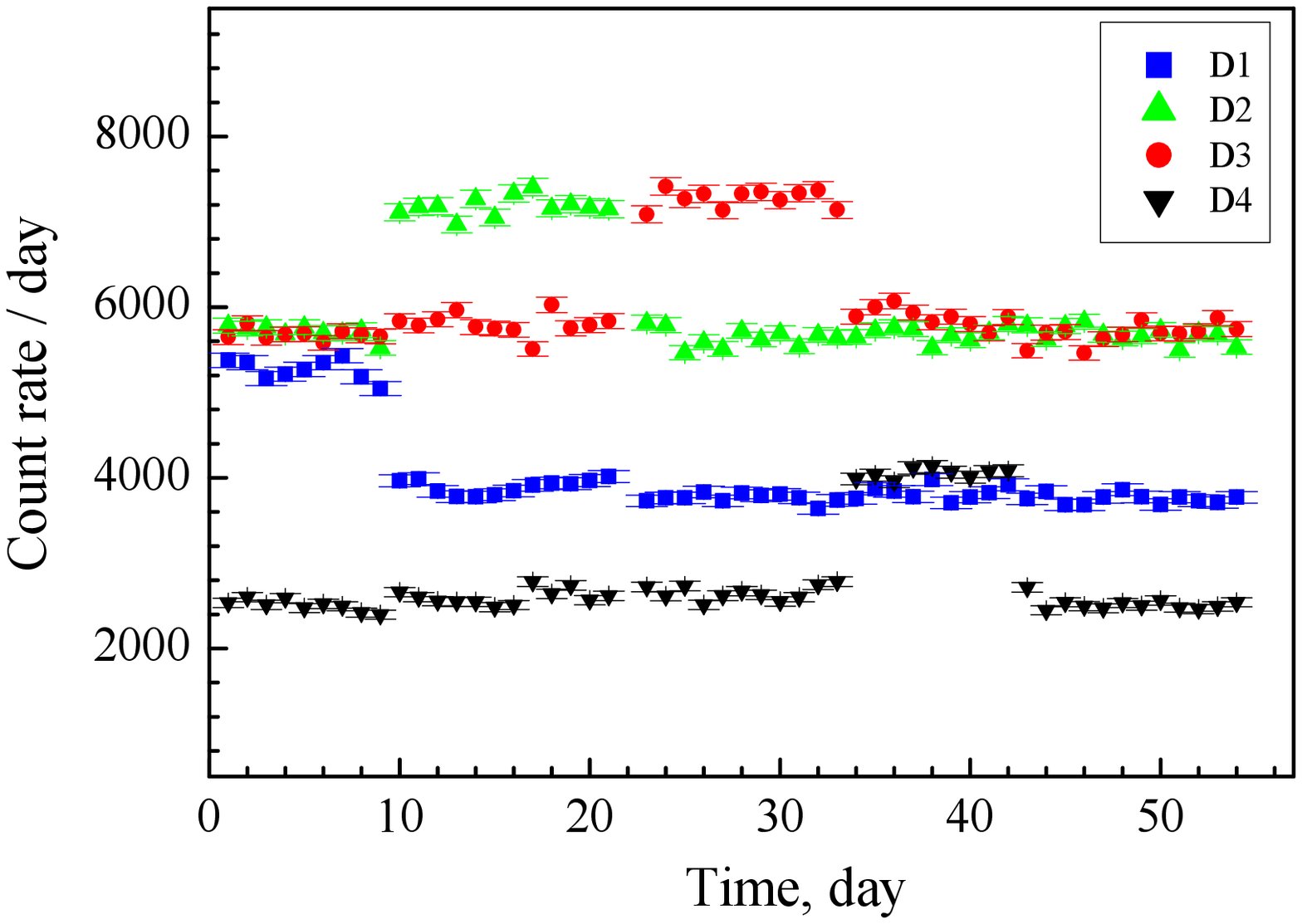}
\caption{\label{fig:calibration}
Count rate of detectors during alternately calibration with $n$-source. Beginning of the calibration corresponds to $\simeq 420$ day in Fig.~\ref{fig:rate}}
\end{center}
\end{figure}

The difference in background count rate of detectors is due to different contamination of $^{210}$Pb in scintillators. Which appears in decay chain of $^{222}$Rn, when during production of scintillators daughter ions of $^{222}$Rn stick to the surface of the scintillator. The half-life of $^{210}$Pb is 22.3~y. It's daughter nuclei $^{210}$Po decays due to $\alpha$-decay inside the scintillator giving a signal similar to that from neutron detection. Assuming that detector D4 has no intrinsic background from $^{210}$Po we may conclude that sum count rate of all detectors from neutrons $\Phi_n \leq 10400$~cts/day at the beginning of calibration, rest $\Phi_r \geq 7800$~cts/day - intrinsic background from $^{210}$Po.

\section{Results}

First stage of the main measurements started 08 August 2012 y and finished 25 September 2013. Second stage started 06 November 2013, with additional elements in the setup: temperature, humidity and pressure sensors. In total, at the moment, we collected data for $\simeq 2.7$ years. The behaviour of the sum count rate of the detectors is presented in Fig.~\ref{fig:rate}.
\begin{figure*}[pt]
\begin{center}
\includegraphics[width=3.5 in, angle=0]{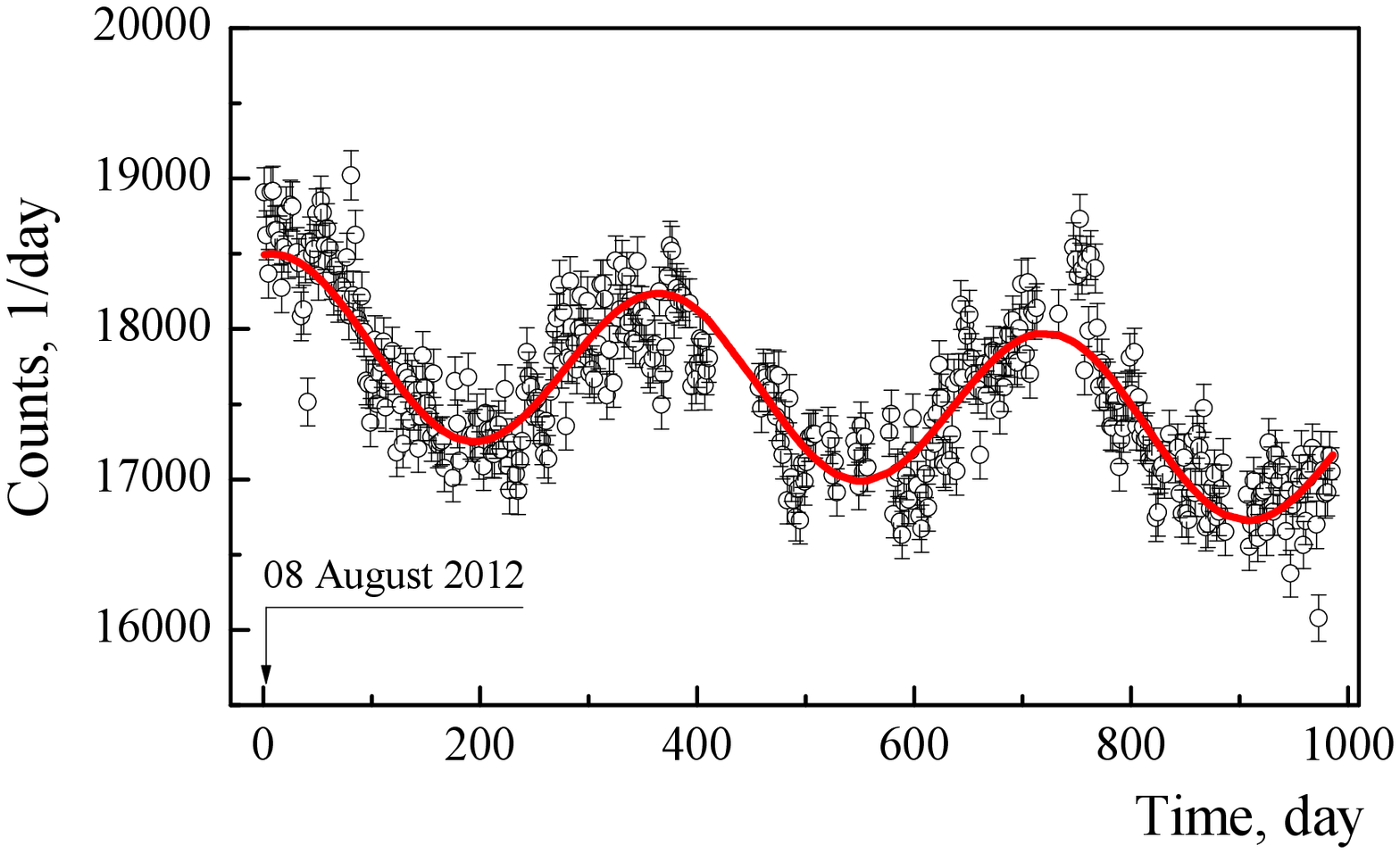}
\caption{\label{fig:rate} Count rate of detectors during $\sim 2.7$ years.}
\end{center}
\end{figure*}
The data, when all conditions were stable are used (ventilation system of tunnel is working well, gates are closed). The count rate is well fitted with cosine function with decreasing base line:

\begin{eqnarray}\label{fit}
 A = a \cdot \texttt{cos} \left ( \frac {2 \cdot \pi \cdot x } {b} + c \right ) + d \cdot x + e,
 \label{DSE}
\end{eqnarray}
where $A$ - count rate [cts/day], $x$ - time [day], $a=556$, $b=356$, $c=-0.208$, $d=-0.736$, $e=17949$. So, the amplitude of neutron flux modulation $\Delta\Phi_n \geq 5$\% 
taking into  account the intrinsic background from $^{210}$Po.

The behaviour of the count rate of the detectors and humidity is very similar to that measured in work {\cite{Stenkin}}. The continuous decrease of count rate is due to decay of $^{210}$Pb. The maximums in count rate corresponds to summer, while minimum - winter. The explanation is, that the count rate depends on absolute humidity of the air in laboratory. Probably, humidity in air works as an additional moderator for neutrons coming from rock. The comparison of sum count rate of the setup with the humidity of the air in lab is presented in Fig.~{\ref{fig:rate_hum_corr}}. It is well seen the strong correlation between count rate and humidity. In Fig.~{\ref{fig:rate_hum_fit}} the dependence of count rate from humidity of the air and fit result by linear function are shown. Thus, additional 1~g/m$^3$ of humidity in air gives a rise in sum count rate of  detectors $\Delta \Phi_n \simeq500$~cts/day. This value is similar to amplitude of the annual modulation of count rate.
\begin{figure*}[pt]
\begin{center}
\includegraphics[width=3.5 in, angle=0]{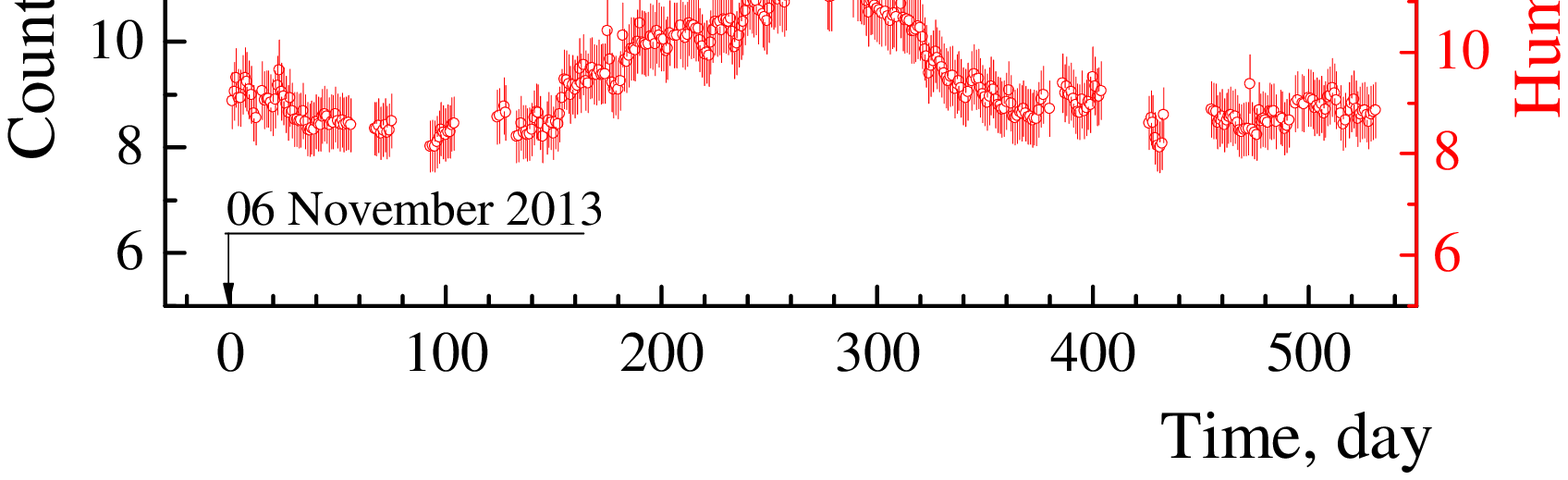}
\caption{\label{fig:rate_hum_corr} Count rate of the detectors and humidity during 1.5 y since 06 November 2013.}
\end{center}
\end{figure*}

It is well seen the strong correlation between count rate and humidity. In Fig.~\ref{fig:rate_hum_fit}
\begin{figure}[pt]
\begin{center}
\includegraphics[width=3.5 in, angle=0]{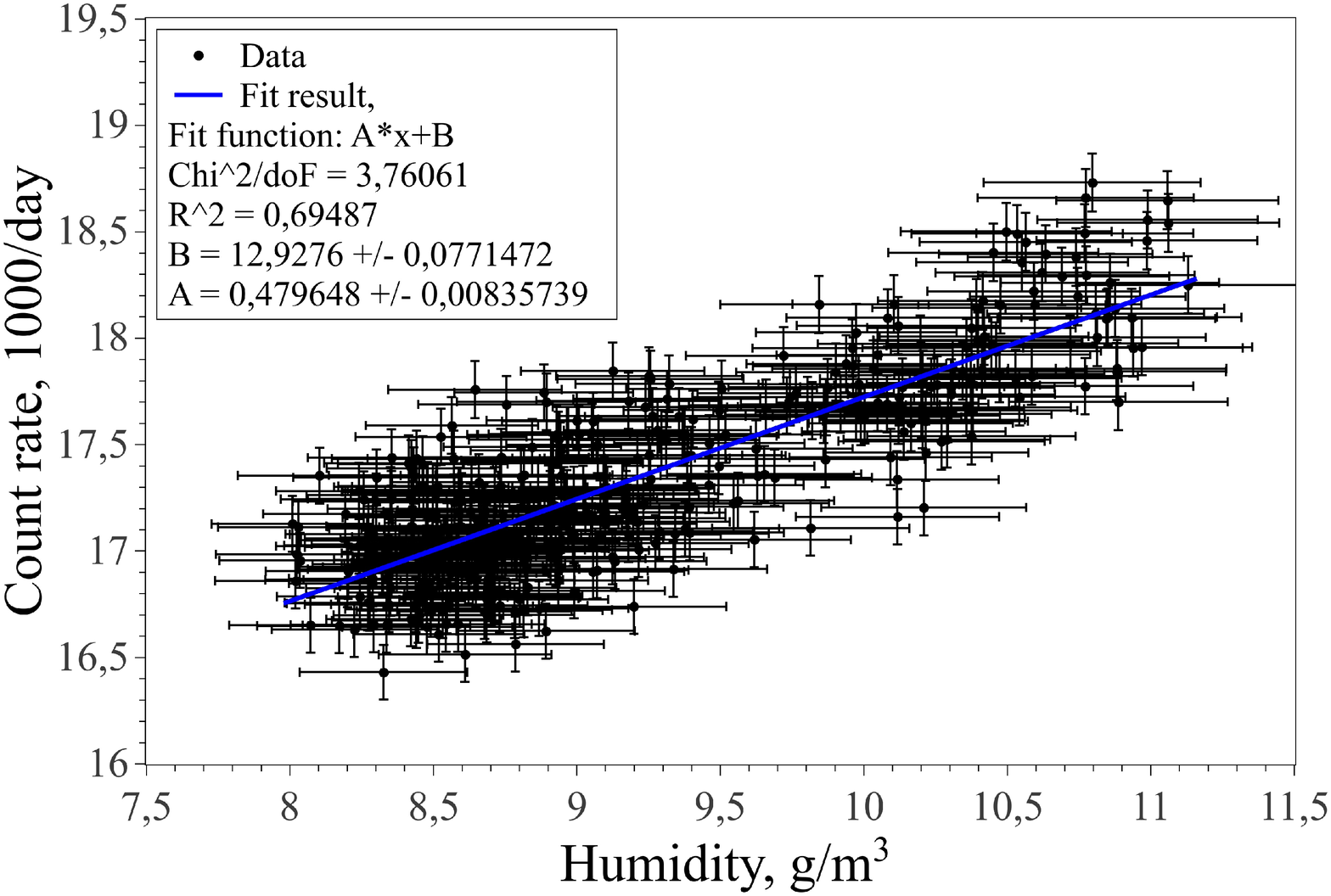}
\caption{\label{fig:rate_hum_fit} Correlation between count rate and humidity.}
\end{center}
\end{figure}
\begin{figure}[pt]
\begin{center}
\includegraphics[width=3.5 in,angle=0]{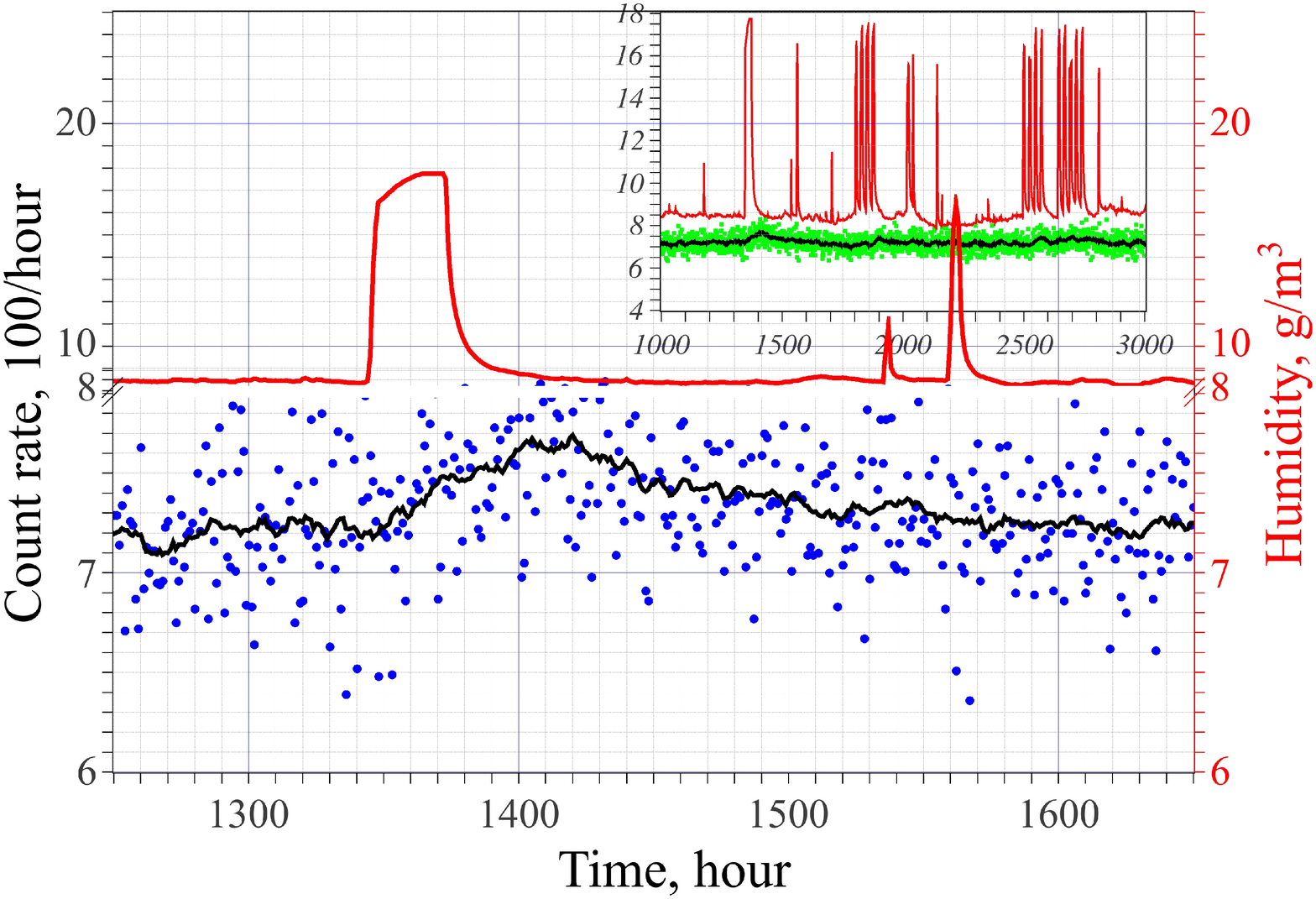}
\caption{\label{fig:hum_eff} Effect of quick rise of humidity on count rate of the detectors.}
\end{center}
\end{figure}
the dependence of count rate from humidity of the air and fit result by linear function are shown. Thus, additional 1~g/m$^3$ of humidity in air gives a rise in sum count rate of  detectors $\Delta \Phi_n \simeq500$~cts/day. This value is similar to amplitude of the annual modulation of count rate.

Another effect seen by our setup is, that in case of quick rise of humidity (for example ventilation of the tunnel is stopped) the count rate start to rise slowly (see Fig.~{\ref{fig:hum_eff}}).
Assuming, that rise in count rate is 30~cts/h during 30~h and that additional humidity of 9~g/m$^3$ should give rise in count rate $\Delta \Phi_n \simeq190$~cts/h, we may conclude, that it takes $\simeq190$~h to get equilibrium. Such delay could be explained by humidification of the rock surface in the lab. It means two possible effect: 1 - better thermalization (moderation) of the neutrons from the bulk rock crossing cracks filled with water (humidity), 2 - higher albedo (reflection, refraction) of the neutrons incoming in to the rock, thus the lab with the humid wall works as a neutron trap. As result total flux of neutrons in the lab become higher.

\section{Conclusion.}
We performed long time measurement the background thermal neutron flux at underground laboratory DULB-4900 of BNO INR RAS, which is located at 3670 m from the entrance to underground facilities. Total time of measurements was equal to 2.7 years. The correlation of the counting rate of thermal neutrons with humidity (high humidity increases the neutron thermalization probability in air) was detected. An observed annual modulation of thermal neutron flux at DULB-4900 exceed 5\% of total neutron flux. Such effects mean, that, for example, experiments looking for WIMPs should be carried out in a labs with walls covered by neutron shield ($\simeq25$ cm of polyethylene $\simeq2$ mm of cadmium sheet, or equivalent).

The measurements are continuing.

The work was made in accordance with INR RAS plan of the Research and Developments.

\end{document}